\documentclass{iopart}
\usepackage{epsfig}
\usepackage{amsfonts}
\usepackage{listings}

\def\text#1{\mathrm{#1}}

\begin{document}

\title{Effect of temperature and velocity on superlubricity}

\author{Joost A van den Ende$^1$, Astrid S de Wijn$^{1,2}$ and Annalisa Fasolino$^1$}
\address{$^1$Radboud University Nijmegen, Institute for Molecules and Materials, Heyendaalseweg 135, 6525 AJ Nijmegen, The Netherlands}
\address{$^2$Department of Physics, Stockholm University, Albanova University Center, SE 106 91 Stockholm, Sweden}

\eads{\mailto{J.vandenEnde@science.ru.nl}, \mailto{A.S.deWijn@science.ru.nl}, \mailto{A.Fasolino@science.ru.nl}}

\begin{abstract}
We study the effects of temperature and sliding velocity on superlubricity in
numerical simulations of the Frenkel-Kontorova model.  We show that resonant
excitations of the phonons in an incommensurate sliding body lead to an
effective friction and to thermal equilibrium with energy distributed over the
internal degrees of freedom.  For finite temperature, the effective friction
can be described well by a viscous damping force, with a damping coefficient
that emerges naturally from the microscopic dynamics.  This damping coefficient
is a non-monotonic function of the sliding velocity which peaks around resonant
velocities and increases with temperature.  
At low velocities, it remains finite and nonzero, indicating the
preservation of superlubricity in the zero-velocity limit.  Finally, we
propose experimental systems in which our results could be verified.
\end{abstract}

\submitto{\JPCM}
\maketitle
\section{Introduction}
Recent developments in surface scanning probes have stimulated a large interest in understanding and using the phenomenon of friction at the nanoscale~\cite{pers,mate}. One of the most fascinating aspects of this field is the possibility of {\it superlubricity}, namely the possibility to slide two surfaces onto each other with vanishingly small friction. Several  experiments report such a low-friction state~\cite{hirano,Dienwiebel,liu}. The crucial ingredient of this phenomenon is the geometrical incommensurability of the sliding crystal faces.
Other aspects like roughness of the sliding surfaces~\cite{rough} or electronic excitations during sliding~\cite{electronic} are not discussed here. The key model to study incommensurate systems is the Frenkel-Kontorova model (FK-model)~\cite{frenkelkontorova}(shown in the top panel of figure~\ref{fig:system}), which describes a chain of atoms with period $a$ on a potential profile with period $b$. The case where the ratio $b/a$ is not a rational number identifies an incommensurate system with no finite global periodicity. 

Aubry has shown that a structural transition occurs at a critical coupling $\lambda_{\text{c}}$ of the chain to the substrate potential~\cite{lc,Aubry}. Below this critical coupling, the static friction vanishes because, while moving, the atoms of the chain always explore all possible positions on the substrate and thereby average out the effective corrugation and resulting friction force. 
One of the empirical laws that describes macroscopic friction states that kinetic friction is lower than static friction~\cite{pers}. Therefore it is important to establish whether vanishing static friction indeed implies vanishing kinetic friction also at the nanoscale. In the paper where they coined the term superlubricity, Shinjo and Hirano~\cite{shinjo} studied the kinetic contribution to friction in the FK model and proposed a phase diagram (bottom panel figure~\ref{fig:system}) with a frictional and a superlubric region as a function of the coupling $\lambda$ and the velocity of the centre of mass of the chain.  In the superlubric regime, once in motion, the chain would slide indefinitely without any kinetic friction but with a recurrent exchange of  kinetic energy between the centre of mass (CM) and a single internal mode related to the periodic modulation due to the substrate. According to their findings, by increasing the velocity of the chain for weak coupling ($\lambda < \lambda_{\text{c}}$, no static friction), namely along a vertical line in the phase diagram of figure~\ref{fig:system}, the system would first acquire friction and then go back to a superlubric state at higher velocities. The proposed behaviour implies a surprising sudden change to a frictional state as soon as the system is set into motion. 

This result has later been criticized~\cite{fkphononconsoli} by showing that the initial superlubric behaviour is destroyed with time, since the recurrent oscillation of the CM gives rise to the resonant parametric excitation of acoustical, long wavelength, vibrations in the chain.
This phenomenon  leads to a high friction regime where all the initial kinetic energy of the CM is eventually converted into internal motion that can be interpreted as  heat. 
The resonant excitation of acoustical modes is however weaker at lower velocities and goes naturally towards the vanishing friction of the static limit. This result agrees well with the fact that most experimental observations of quasi-vanishing friction have been at the extremely low velocities of atomic force microscopes (AFM), of the order of $\mu$m/s. Indeed, the term superlubricity is currently used to indicate in general a low friction sliding rather than in the original meaning of reference~\cite{shinjo}.

Here we study how superlubricity is affected by temperature and velocity. First we show that the dynamical mechanism identified in reference~\cite{fkphononconsoli} leads to a system in thermal equilibrium that satisfies energy equipartition over the internal vibrations, yielding a well defined temperature.
We then show that at finite temperature the system displays a damped dynamics that allows to calculate the damping coefficient from the microscopic description rather than by assuming its presence phenomenologically. We find that this damping is not a monotonic function of the sliding velocity, as it peaks close to specific phonon resonances.  Away from these resonances and particularly in the limit of low velocity, the effective damping is finite, and thus at low velocities the friction vanishes, preserving superlubricity. 

The paper is organized as follows. In section \ref{mod} we describe the model and its dynamics, in section \ref{diss} we review how kinetic friction arises and show how the system evolves into thermal equilibrium. In section \ref{visc} we show that, at thermal equilibrium, friction can be described as a viscous damping that can be calculated from the model. We also compare our results to damped driven models~\cite{StrunzElmer}. In section \ref{est} we suggest experimental systems that could confirm the results presented in this paper and in section \ref{sum} we summarize our findings.

\section{Model: geometry and dynamics}
\label{mod}
\begin{figure}
\epsfig{figure=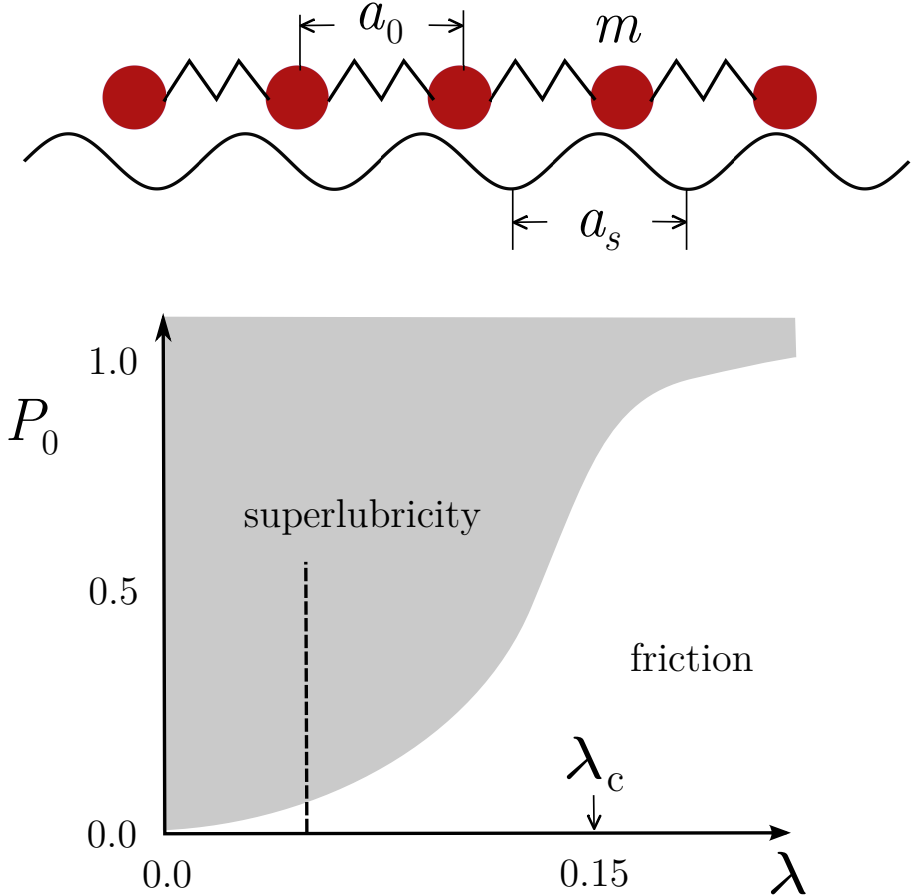,width=12cm}
\caption{
\label{fig:system}
Top panel: a schematic overview of the Frenkel-Kontorova (FK) model.
A chain of harmonically coupled particles is subjected to a periodic potential.
Bottom panel: a sketch of the proposed phase diagram by Shinjo and Hirano (adapted from reference~\cite{shinjo}). As the initial CM velocity is increased along the dashed line, the system supposedly first undergoes a transition from zero static friction ($\lambda < \lambda_{\text{c}}$) to kinetic friction and then to a superlubric state. Our results do not support this picture.
}
\end{figure}

The FK-model~\cite{frenkelkontorova} consists of a harmonic linear chain of particles of mass $m$ with period $a_0$, subjected to an external periodic potential with period $a_{\text{s}}$ and amplitude $U_0/2 \pi$. The spring constant between the particles is $K$. 
The ratio of the depth of the potential to the spring constant of the chain defines a coupling parameter: $\lambda \equiv U_0/(K a_{\text{s}}^2)$. 
In the context of friction the particles represent the atoms of a sliding body and the potential represents a rigid solid substrate. The model is depicted schematically in the top panel of figure~\ref{fig:system}.

If the ratio of the parameters $a_0/a_{\text{s}}$ is an irrational number, the system is called incommensurate, otherwise the system is called commensurate.
The incommensurate case is the most interesting, because for two arbitrary surfaces in contact a common periodicity is not likely.

The ground state is determined by the competition of the harmonic interaction between the particles and the interaction of the particles with the periodic substrate, which is controlled by $\lambda$.
For incommensurate systems there exists a structural phase transition~\cite{lc,Aubry}: for $\lambda < \lambda_{\text{c}}$ the system is in a so called floating phase and there is no static friction. If $\lambda > \lambda_{\text{c}}$ the system is in the pinned phase and exhibits static friction.  The value of $\lambda_{\text{c}}$ is largest when the ratio between the length scales is equal to the golden mean, $\tau_{\text{g}}$, $a_0/a_{\text{s}}=\tau_{\text{g}}=(1+\sqrt{5})/2=1.618033989 \ldots$~\cite{greene}, for which $\lambda_{\text{c}} =0.1546 \ldots$~\cite{mac}. In the numerical implementation, one may approximate the golden mean by a ratio of successive Fibonacci numbers, $\tau_{\text{g}}=\lim_{n \rightarrow \infty} F_{n+1}/F_n$. Here, all calculations  are done for $n=12$, i.e. $N=144$ particles are distributed onto $233$ periods of the substrate potential.

The model can be enriched by considering the dynamics of the particles~\cite{shinjo,fkphononconsoli}.
We include the kinetic energy of the particles by defining  $\tau= t/\tau_0$, where $t$ is the time and $\tau_0 \equiv \sqrt {m/K}$. If we express all energies in units of $K a_{\text{s}}^2$,  times in units of $\tau_0$ and lengths in units of  $a_{\text{s}}$, the Hamiltonian of the system becomes:

\begin{equation}
\label{FK}
\mathcal{H}= \sum_{i=1}^{N} \left[ \frac{1}{2} \left( \frac{\rmd u_i}{\rmd \tau} \right)^2 + \frac{1}{2}(u_{i+1}-u_i-a_0)^2 + \frac{\lambda}{2\pi} \left(1 - \cos \left(2 \pi u_i \right) \right) \right]~,
\end{equation}

where $u_i$ is the position of particle $i$ of the chain with the periodic boundary condition $u_{i+N}=u_i$.
The resulting equations of motion can be written as a dynamical system of $2N$ first order ordinary differential equations:
\begin{eqnarray}
\frac{\rmd u_i}{\rmd \tau}=  y_i \label{eqm1}, \\
\frac{\rmd y_i}{\rmd \tau}=  u_{i+1}+u_{i-1}-2u_i- \lambda \sin (2 \pi u_i)~.
\label{eqm2}
\end{eqnarray}

We solve these equations numerically by using the fourth order Runge-Kutta (RK) method with
 a time step of $\tau_0/150$ and calculate also 
the position $Q=(1/N)\sum_{i=1}^{N}u_i$ and velocity $P=(1/N)\sum_{i=1}^{N}y_i$ of the CM.
As the system is Hamiltonian (no heuristic damping) the total energy is conserved.
Nevertheless, energy can be transferred from the centre of mass to the internal degrees of freedom, leading to the arrest of the chain in time. This effect can be interpreted as an effective friction.

If there is no coupling between the chain and the substrate, $\lambda=0$, the dispersion relation of the phonon modes is the one of a harmonic chain, $\omega_k=2|\sin\left(k/2\right)|$, with $\omega_k$ the frequency of the phonons and $k=2\pi i/N$ the wavevectors in which $i \in (-N/2,N/2]$.
When all the particles move with the same velocity $p_0$, the atoms slide over the periodic potential with the washboard frequency $\Omega=2 \pi p_0$.
Strong resonances occur if $\Omega$ is close to the frequency of the phonon mode with the wavevector $q = 2 \pi a_0/a_{\text{s}}$ or to its harmonics $nq$ of multiples of the wavevector $q$. If $\lambda$ is small, these frequencies can be approximated by the frequencies $\omega_{n\text{q}}$ of the phonons of the harmonic chain.
The resonance phenomenon therefore occurs for CM velocities~\cite{StrunzElmer,fkphononconsoli}:

\begin{equation}
\label{res}
np_0\sim\frac{\omega_{n\text{q}}}{2 \pi}  \text{\quad} n \in \mathbb{N}  \text{ \quad} n \geq 1.
\end{equation} 

When these resonances occur, after an initial recurrent energy exchange~\cite{shinjo,fkphononconsoli}, the CM translational kinetic energy is eventually totally transformed into internal kinetic and potential energy of vibration in the chain. This energy transfer occurs because the CM motion with velocity $P$ induces a modulation with the washboard frequency $\Omega$ that leads to parametric resonances with exponential growth of phonon modes with wavevector $k$ 
whenever~\cite{StrunzElmer,fkphononconsoli,fkphononconsoli2} 
\begin{equation}
\Omega\simeq \frac{\omega(k)+\omega(mq-k)}{m}
\end{equation}
for some integer $m$.
From the above relation, one can derive a maximum value $P_{max}= \Omega_{max}/(2\pi) =4|\cos(\pi \tau_g/2)|/(2\pi) \approx 0.5254$. Above this value, the transfer of energy from the CM to the phonons is blocked and the motion would be really superlubric. 

Lastly we note that, although the total energy is conserved, entropy increases when the energy is
dispersed over the large number of internal degrees of freedom of the chain.
Finite systems can show temporary decreases in the entropy, and, in principle,
due to Poincar\'e recurrence, the system could return close to the initial
state with all energy in the CM momentum. In
practice, however, for systems with a sufficiently large number of internal
degrees of freedom, such a large decrease in entropy is highly improbable and
occurs only on extremely long time scales.  Indeed, we have never observed this reversal in the length of our simulations.

\section{Dissipation leading to thermal equilibrium}
\label{diss} 

In this section, we show the evolution of the CM velocity $P$ for initial velocities close to the first four resonances ${\omega_{n\text{q}}}$ and show how this process leads to  thermal equilibrium. We take $\lambda=0.05 \approx \lambda_{\text{c}}/3$ for which there is no static friction, namely we consider values of $P$ on the vertical dashed line in figure \ref{fig:system}. For the initial positions of the particles we take a ground state of the system, calculated by minimizing the potential energy. We give all the particles the same initial velocity in order to have the CM velocity $P$ equal to the velocity corresponding to the first four resonances (table~\ref{resinfo}).

\begin{figure}
\epsfig{figure=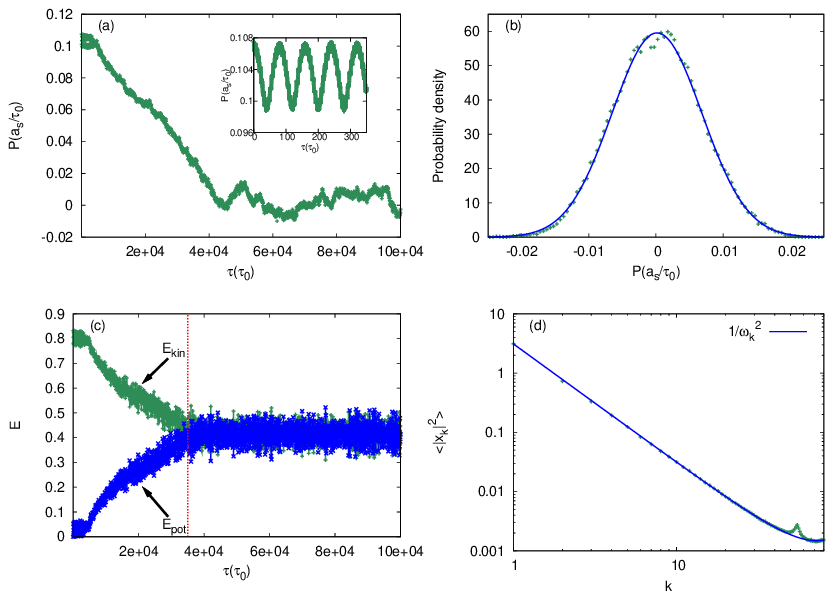,width=12cm}
\caption{     
\label{fig:equilibration}
Time evolution of the system for the second resonance. (a) $P$ as a function of $\tau$ in the region in which $P$ decays with time. The inset shows the initial recurrent mechanism. (b) Probability density function of $P$ from data of the $P$-$\tau$ evolution (points) and the Gaussian fit of this function (line). (c) $E$ as a function of $\tau$ for the second order of resonance. The total kinetic energy and the increase in potential energy relative to the minimum configuration are shown. The dashed vertical line (red) is the time at which equipartition is reached. (d) The amplitudes of the phonon modes $<x_k^2>$ as a function of mode number $k$ for the second order of resonance. The data points are plotted in green and the fit according to $1/\omega_k^2$ is in blue. The deviation of the $1/\omega_k^2$ behaviour around $k=55=-q$ is due to the approximated incommensurability of the system. For the time ranges used for averaging  in (b) and (d), see Table \ref{resinfo}.
}
\end{figure}

In figure~\ref{fig:equilibration} we show the time evolution $P(\tau)$ for the second resonance. We see an initial oscillatory phase, enlarged in the inset of figure \ref{fig:equilibration}(a), where $P$ periodically goes back to the initial value. After some time this recurrence mechanism breaks down and $P$ decays with time.  This effect, the onset of friction, was not reported in reference~\cite{shinjo}, possibly, because of the use of a too small number of particles or too short integration times~\cite{fkphononconsoli}. The mechanism for this behaviour, parametric excitation of acoustic phonons, has been analyzed in reference~\cite{fkphononconsoli}. A similar behaviour has also been found for sliding concentric nanotubes~\cite{transphonon}. 

When the decay has finished, $P$ jiggles around 0.  From figure~\ref{fig:equilibration}(c) we see that the CM kinetic energy has been transformed in equal amounts of kinetic and potential energy as expected classically for vibrations, a first indication that the system has reached thermal equilibrium. 
For the other resonances, the evolution of $P$ in time is  qualitatively similar but with different time scales: the higher the order of resonance the longer the time before the start of the decay of $P$. For the fourth resonance, we had to use a slightly perturbed ground state to speed up this process. The increasingly long time it takes for the parametric growth of the phonon excitations to lead to friction makes it very difficult to study numerically the low velocity limit. 
We cannot rule out superlubric motion at very low velocities where the parametric growth of internal vibrations could occur either on time scales too long for the simulations or not occur at all.  Notice that these results do not support the phase diagram~\cite{shinjo} reported  in figure~\ref{fig:system}.

If, after the decay of $P$, the system has reached thermal equilibrium, equipartition should apply, namely the initial CM kinetic energy should be distributed over the $N$ oscillators, each having $k_{\text{B}}T_{\text{eq}}$.  We use two independent methods to ascertain if this is the case.  

At thermal equilibrium, the probability density function of $P(\tau)$ should be described by a Maxwell-Boltzmann (MB) distribution. In one dimension, the MB distribution coincides with a Gaussian distribution around $0$, with variance $\sigma^2=k_{\text{B}}T_{\text{eq}}^{\text{CM}}/M$, where $M=Nm$ is the mass of the chain, $k_{\text{B}}$ is the Boltzmann constant and $T_{\text{eq}}^{\text{CM}}$ the equilibrium temperature. We calculate the probability density function from a normalized histogram of $P$, divided in 100 uniform bins, in the time interval in which $P$ jiggles around $0$. The result, shown  for the second resonance  in figure~\ref{fig:equilibration}(b) is indeed described well by a Gaussian  with mean $\mu=1.5 \times 10^{-4}~ a_{\text{s}}/\tau_0$ and variance $\sigma^2=4.5\times10^{-5}~a_{\text{s}}^2/\tau_0^2$.  The estimated values for the equilibrium temperature are given in table~\ref{resinfo}.

Another estimate for the temperature can be given by the amplitudes, $x_k$, of the phonon modes~\cite{fkphononconsoli2}: $\sum_{k=1}^{N} \frac{1}{2} m \omega_k^2 \langle x_k^2 \rangle = \frac{1}{2} k_{\text{B}} T_{\text{eq}}^{\text{phon}}$, where the brackets denote a time average and $T_{\text{eq}}^{\text{phon}}$ is the equilibrium temperature corresponding to the distribution of the phonon amplitudes. In figure~\ref{fig:equilibration}(d) we show, for the second resonance, that the amplitudes of the phonon modes are proportional to $1/\omega_k^2$ as expected at equilibrium.

The behaviour of the distribution of $P$ and the phonon modes is qualitatively the same for the other resonances. In table~\ref{resinfo} we  compare the temperatures $T_{\text{eq}}$  with the values obtained from the distribution of $P$, $T_{\text{eq}}^{\text{CM}}$, and from the phonon amplitudes, $T_{\text{eq}}^{\text{phon}}$.  In order to define high and low temperature regimes, all temperatures are given in units of the value of $\lambda$ used in this work, 0.05. In this way we can distinguish high ($T \approx 1~\lambda$)  and low temperature ($T<<1~ \lambda$) regimes. In the next section, we also study the generic value $T=0.16~ \lambda$, between first and second resonance, for which the distribution of $P$ was checked and found to be a Gaussian.
The agreement between the different estimates demonstrates that the system has reached thermal equilibrium in the studied range of velocities around the first four resonances.         

\begin{table}[!h]
\caption{For each resonance $n$ we give the initial velocity per particle $p_0$. $\tau$-range gives the time range in which the probability density function of $P$ and the averages of the phonon amplitudes have been calculated. The absolute value of the averages ($\mu$) of the Gaussian fits of the histograms of $P$ is given in the fourth column in $10^{-4}a_{\text{s}}/\tau_0$. They are a measure of the deviation of the system from a perfect thermal equilibrium. The last three columns are the temperatures obtained from the initial kinetic energy ($T_{\text{eq}}$), the distribution of $P$ ($T_{\text{eq}}^{\text{CM}}$) and the phonon modes ($T_{\text{eq}}^{\text{phon}}$).} 
\begin{tabular}{|p{0.4cm}|p{1.4cm}|p{2.4cm}|p{1.2cm}|p{0.9cm}p{2.0cm}p{2.0cm}|}
\hline
$n$ & $p_0$($a_{\text{s}}/\tau_0$) & $\tau$-range ($\tau_0$) & $|\mu|$& $T_{\text{eq}}(\lambda)$ & $T_{\text{eq}}^{\text{CM}}(\lambda)$  & $T_{\text{eq}}^{\text{phon}}(\lambda)$ \\
\hline
1 & 0.2966 & $(5-405)\times10^4$  & $6.5  \pm 2$   & 0.88  & $0.89 \pm 0.01$   &$0.91 \pm 0.04$    \\
2 & 0.1075 & $(25-425)\times10^4$ & $1.5  \pm 0.7$ & 0.12 & $0.13 \pm 0.01$ & $0.13 \pm 0.01$  \\
3 & 0.04692& $(4-16)\times 10^6$  & $3.1   \pm 1  $& 0.022 & $0.026 \pm 0.002$ & $0.030 \pm 0.004$   \\
4 & 0.07927& $(4-8)\times 10^6$   & $2.0 \pm 0.7$  & 0.063 & $0.055 \pm 0.002$ & $0.074 \pm 0.004$   \\
\hline
\end{tabular}
\label{resinfo}
\end{table}

\section{Emergence of viscous damping}
\label{visc} 

After the system has reached thermal equilibrium, as described in the previous section, we give the CM an initial velocity $P(\tau=0)$ and study how it evolves with time.
If the friction is purely viscous, $F_{\text{fric}}=-\eta M P$, as is often assumed in the study of frictional dynamics~\cite{StrunzElmer,fil,mus}, the sliding velocity decays exponentially with time with a time scale determined by the friction coefficient $\eta$. 
This type of viscous behaviour can be expected on general theoretical grounds for many-particle systems~\cite{vankampenadiabaticelim,jarzynskitss}.
 
\begin{figure}
\epsfig{figure=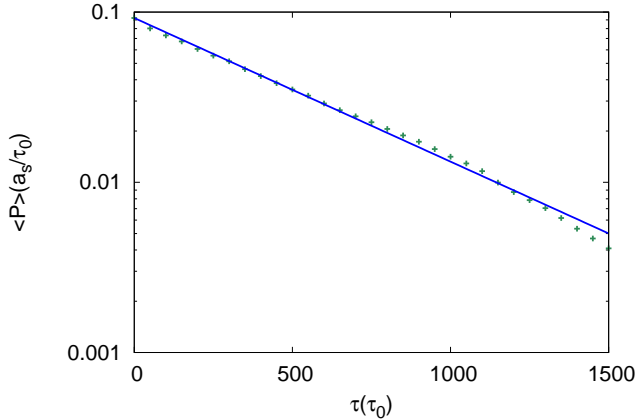,angle=270,width=8.6cm}
\caption{
\label{fig:exponentialdecay}
{Illustration of the way to calculate $\eta$. The evolution of $\langle P(\tau)\rangle$ with respect to time after the addition of a velocity $p^+=0.09$ to all the particles for $T=0.88$ $\lambda$,
averaged over 500 initial configurations chosen with a spacing in time of $2000~\tau_0$ from the trajectory starting from the first resonance, with $T=0.88~\lambda$.
The fit is an exponential function, $\langle P(\tau) \rangle(a_{\text{s}}/\tau_0) =0.0923 \exp(-\eta \tau)$ with $\eta=0.00194(1/\tau_0)$.
}
}
\end{figure}

We start from initial conditions obtained by choosing many (typically hundred) configurations of the system in thermal equilibrium at temperature $T_{\text{eq}}$ in the $\tau$-range given in table~\ref{resinfo}.
We increase the CM velocity $P$ by adding a uniform initial sliding velocity $p^+$ to all the particles.
The systems are then evolved with time and the time dependence of the averaged CM velocity $\langle P(\tau) \rangle$ is obtained by averaging over all initial configurations.
This averaging is necessary, as, in such small systems, fluctuations are large.
In figure~\ref{fig:exponentialdecay} we show $\langle P(\tau) \rangle$ for  $p^+=0.09 ~ a_{\text{s}}/\tau_0$, averaged over 500 configurations at $T=0.88 ~ \lambda$.
The average CM velocity decays exponentially, $\langle P(\tau) \rangle(a_{\text{s}}/\tau_0)=0.0923 \exp (- \eta \tau)$ with a friction coefficient  $\eta=1.94 \times 10^{-3}(1/\tau_0)$.

In this system, it is the microscopic dynamics that give rise to the friction and determine the value of the friction coefficient, without the need for imposing a phenomenological friction term.
A related result for damping of phonon modes in relation to thermostats in crystalline materials was obtained 
and tested in the context of friction~\cite{BenassiPRL}.

The time range used for the calculation of $\eta$ needs to be not too short because of the reliability of the fit, nor too long because of the change of temperature of the system which occurs during the time range. We choose for the high temperature case, the time range $\tau=0-500~\tau_0$ and for the low temperature case, the time range $\tau=0-1000~\tau_0$, to determine $\eta$ and its error bar. Therefore, we do not consider the effect of different time ranges for the fit, although we have checked that the behaviour of $\eta$ remains qualitatively the same.  

\begin{figure}
\epsfig{figure=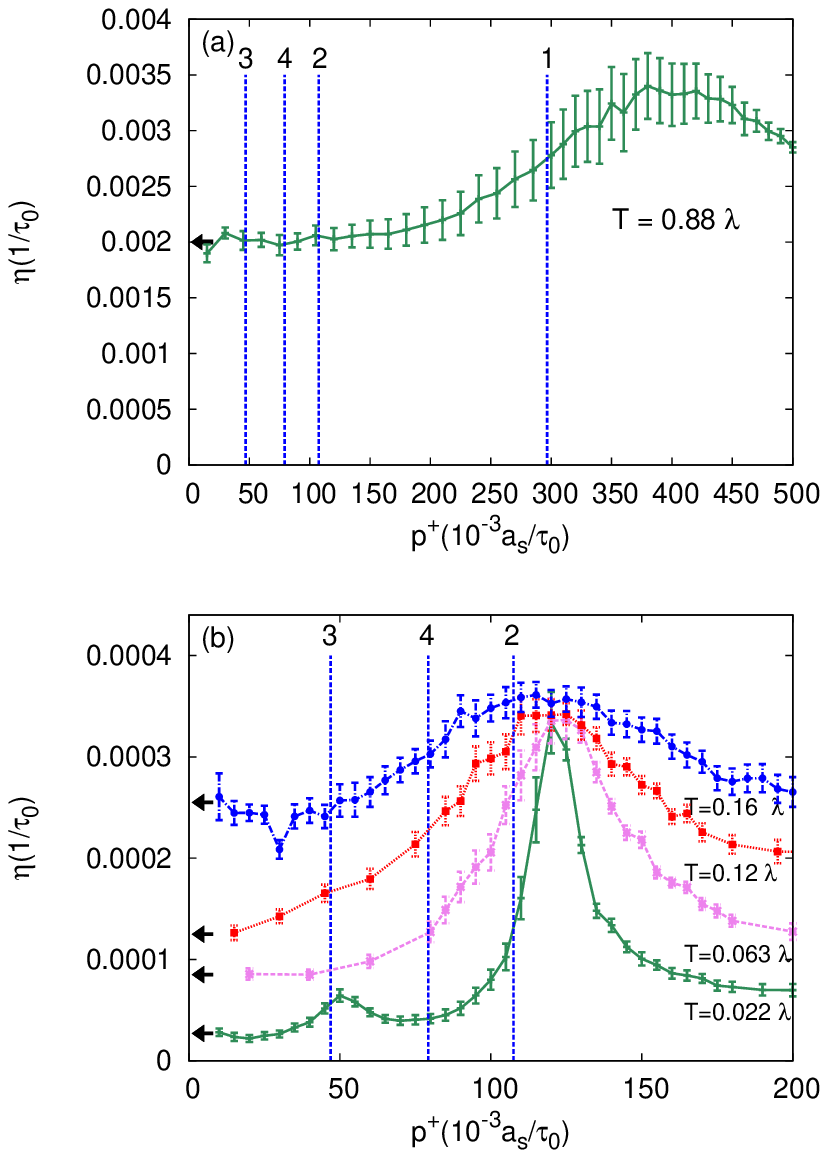,width=8.6cm}
\caption{
\label{fig:effectivedamping}
{$\eta$ as a function of $p^+$ for different temperatures, (a) the high temperature case, $T=0.88$ $\lambda$, and (b) the low temperature cases.
The vertical dashed lines are the values of $p^+$ corresponding to the first four resonances (equation~(\ref{res})).
The peak resulting from the resonance broadens at higher temperatures.
Each friction coefficient was obtained from averaging over at least 100 initial configurations separated $2000~\tau_0$ in time. $\eta$ has been obtained from $\tau=0-500 ~ \tau_0$ in (a) and from $\tau=0-1000 ~ \tau_0$ in (b).}
}
\end{figure}

The behaviour of $\eta$ as a function of sliding velocity is shown in figure~\ref{fig:effectivedamping}.
For all $T$, in the limit $p^+\rightarrow 0$, $\eta$ becomes constant (as indicated by the arrows in figure~\ref{fig:effectivedamping}), implying that the viscous friction, $F_{\text{fric}}$, goes to zero and superlubricity is conserved. The same conclusion has been drawn in \cite{jinesh2008} for the Tomlinson model. 
The friction coefficient $\eta$ grows with temperature. This is not surprising, because thermal fluctuations distort the incommensurate contact between the surfaces. 
It is known that friction decreases with temperature, particularly in the stick slip regime~\cite{temp,jinesh2008} where the `slips' from one minimum of the potential energy to the other can be activated by thermal energy. The attempt frequency of these processes is determined by the damping $\eta$ that is usually taken to be temperature independent.   

The vertical dashed lines in figure~\ref{fig:effectivedamping} are the velocities corresponding to the first four resonances of the system (equation~(\ref{res})). For the high temperature case in figure~\ref{fig:effectivedamping}(a), the friction coefficient  has a peak at frequencies just above the first resonance, $\omega_{q}$ and no pronounced features at the lower three resonances. For the low temperature case, figure~\ref{fig:effectivedamping}(b), the influence of the lower resonances on the friction characteristics becomes apparent. Again the peaks in $\eta$ are shifted to the right. In addition, the influence of temperature on the broadening of the peaks is visible, the higher the temperature, the broader the peak. 
For the lowest studied resonance, the fourth, there is no clear peak of $\eta$  at any of the studied temperatures, whereas for the third resonance it can be seen at the lowest studied temperature. Nevertheless, at sliding velocities near the resonances, superlubricity is suppressed for all temperatures. 

In section~\ref{mod} we have shown that at $T=0$ there is a maximum resonance velocity, $P_{max} \approx 0.5254$.
For initial velocities sufficiently far above this value, there can be no effective damping at $T=0$.
At finite temperature, thermal fluctuations distort the incommensurate geometry, and nonzero effective damping can still exist even for high velocities.  We have indeed observed a slow decay of velocity
in our simulations at $p^+=0.75$ and $T=0.88\lambda$.

Our results can be compared to the  velocity-force characteristics calculated by Strunz and Elmer~\cite{StrunzElmer} for a driven and damped FK-model at $T=0$. The inherent friction in their system, calculated by subtracting the  imposed damping term from the total force, shows an effect of the resonances similar to our results of figure~\ref{fig:effectivedamping}.  A difference between our results and those of Strunz and Elmer is the location of the peaks in the friction characteristics. In reference~\cite{StrunzElmer} these peaks are situated at the values of the resonances calculated from equation~(\ref{res}), while in our situation the peaks are shifted towards higher values of the sliding velocity.
A possible explanation for this shift towards the right in our undamped case, is the influence of the use of the approximation $\lambda=0$ for the resonances equation~(\ref{res}). Indeed, for smaller coupling, $\lambda=0.005$ and $T=8.8~\lambda$, we have observed a smaller shift for the first resonance. 
 
Also Strunz and Elmer used this approximation, but they studied the so called uniform sliding state in a damped system, whereas we solve the full dynamics. Our results confirm the influence of the resonances on the friction found in reference~\cite{StrunzElmer}, and show that this aspect of the model is not limited to $T=0$ nor to  uniform sliding states of a damped driven system.

\section{Estimate of parameters}
\label{est}
Our 1D model is highly simplified and focuses on the geometrical incommensurability which is essential for superlubricity. In real systems also other aspects like roughness~\cite{rough} and electronic friction~\cite{electronic,nano,agxe1} may play a role. Nevertheless, we attempt to find possible experimental situations in which our results could be confirmed. An example of a well studied system with incommensurate contacting surfaces is a layer of Xe atoms sliding on an Ag(111) substrate~\cite{agxe1,agxe2,agxe3}.  
The ratio between the lattice parameters of the Xe monolayer and the substrate is $a_0/a_{\text{s}}=(4.55 \times 10^{-10} \text{m} /2.892 \times 10^{-10} \text{m})  = 1.57 \approx \tau_{\text{g}}$.
The mass of the Xe atoms is $m = 2.16 \times 10^{-25} $~kg.
The spring constant between the Xe atoms can be obtained from a Taylor expansion of the Lennard-Jones potential with parameters for Xe-Xe around its minimum~\cite{agxe1}, $K=1.103~\text{J}/\text{m}^2$.
Consequently, the typical scale of the velocity is $a_{\text{s}}/\tau_0=6.5 \times 10^2$~m/s.
In a similar way $U_0$ is estimated to be in the range $1 - 5 \times 10^{-21}$~J~\cite{agxe1,difbar}, which gives   $ 0.01 < \lambda < 0.05 < \lambda_c$ close to the value of $\lambda$ we used in the simulations.  

The velocity where the effective friction becomes a strongly nonlinear function of velocity is rather high, order of magnitudes too high for AFM and at the limit of reachable velocities for QCM. Therefore it is important to realize that in the low velocity limit, friction grows smoothly with velocity from  the vanishing static friction, contrary to the Shinjo-Hirano  model of figure \ref{fig:system}. It would be interesting however to explore also the transition from low viscous friction to high friction in correspondence of the resonances described before. The friction characteristics for the third resonance could possibly be seen at sliding velocities of approximately 30 m/s, for temperature in the range of about 8 K for Xe sliding on Ag(111).  The second resonance, at around 70 m/s, would be visible up to higher temperatures, at least around 50 K.

Another interesting experimental setting for the FK model is in colloidal suspensions, where atomic-scale systems can be reproduced on larger scales, with colloidal particles playing the role of atoms.
Recently, the static FK model was studied experimentally in such systems~\cite{colloidFK}.
However, as colloidal crystals have over-damped decay of phonon modes~\cite{colloidphonon}, the present results cannot easily be scaled up in the same way.

\section{Summary} 
\label{sum}
We have numerically calculated effective friction coefficients as a
function of temperature and velocity in the FK-model without any driving forces or
phenomenological damping terms.
We have critically reviewed previous findings and shown how initial kinetic
energy of the centre of mass is converted into heat by excitation of the
internal motion of the chain, bringing the system to thermal equilibrium.
We have then studied the temperature and velocity dependence of friction
and found that it can be described
as an effective viscous damping.   The friction coefficient however, shows a
peculiar velocity dependence with peaks at sliding velocities close to
resonances between the phonons of the chain and wavevectors related to the
modulation induced by the incommensurate periodic potential. We argue that
these peaks in the friction coefficient  could be observed
experimentally. Moreover,we find that for incommensurate contacts, the effective damping  increases with temperature.

Despite the fact that  both sliding velocity and temperature can suppress
superlubricity in incommensurate systems, the fact that the
friction coefficient remains finite at low velocities, makes that 
superlubricity is preserved in the important limit of very low velocity
probed by AFM experiments.

\section*{Acknowledgements}
ASdW's work was financially supported by a Veni grant of Netherlands
Organisation for Scientific Research (NWO) and by an Unga Forskare grant from
the Swedish Research Council. This work is part of the research program of the Stichting voor Fundamenteel Onderzoek der Materie (FOM), which is financially supported by the Nederlandse Organisatie voor Wetenschappelijk Onderzoek (NWO).

\section*{References}
\providecommand{\newblock}{}

\end{document}